\documentclass[10pt,twocolumn]{revtex4-1}
\usepackage[margin=0.75in]{geometry}

\usepackage[dvipsnames]{xcolor}
\usepackage{graphicx}
\usepackage{dcolumn}
\usepackage{bm}
\usepackage{stmaryrd}
\usepackage[sort&compress]{natbib}
\usepackage{color}

\newcommand{\comment}[1]{}

\newtheorem{problem}{Problem}

\newtheorem{note}{Note}

\newcommand{\bp}{\begin{problem}}
\newcommand{\ep}{\end{problem}}
\newcommand{\bn}{\begin{note}}
\newcommand{\en}{\end{note}}

\newcommand{\be}{\begin{equation}}
\newcommand{\ee}{\end{equation}}
\newcommand{\bal}{\begin{align}}
\newcommand{\eal}{\end{align}}

\graphicspath{{Figures/}}

\renewcommand{\ee}{\mathbf{e}}
\newcommand{\xx}{\mathbf{x}}

\newcommand{\nn}{\mathbf{n}}

\newcommand{\ff}{\mathbf{f}}
\newcommand{\FF}{\mathbf{F}}
\newcommand{\uu}{\mathbf{u}}

\usepackage{pgfplots}
\usetikzlibrary{matrix}
\usepgfplotslibrary{groupplots}
\usepgfplotslibrary{patchplots}
\usepackage{color}
\usepackage{array}
\usepackage{amsmath,amsfonts,bm}
\usepackage{graphicx,epsfig}

\newcolumntype{C}[1]{>{\centering\let\newline\\\arraybackslash\hspace{0pt}}m{#1}}

\usepackage{booktabs}
\usepackage{physics}
\usepackage{amsmath,amssymb}
\usepackage{bm}

\begin{document}
%
%
\title{Soft-Lubrication Drainage and Rupture in Particle-Driven Vesicles}

\author{Yuan-Nan Young$^{1}$}
\email[Corresponding authors:]{yyoung@njit.edu, has@princeton.edu}
\author{Bryan Quaife$^{2}$}
\author{Herve Nganguia$^{3}$}
\author{On Shun Pak$^{4}$}
\author{Jie Feng$^{5}$}
\author{Howard A. Stone$^{6*}$}
\affiliation{$^{1}$Department of Mathematical Sciences, New Jersey Institute of Technology, Newark, New Jersey, 07102, USA\\
$^2$Department of Scientific Computing, Florida State
 University, Tallahassee, Florida, 32306, USA\\ 
$^3$Department of Mathematics, Towson University, Baltimore, Maryland, 21252, USA\\
$^4$Department of Mechanical Engineering, Santa Clara University, Santa Clara, California, 95053, USA\\
$^5$Department of Mechanical Sciences and Engineering, University of Illinois Urbana-Champaign, Urbana, Illinois, 61801, USA\\
$^6$Department of Mechanical and Aerospace Engineering, Princeton University, Princeton, New Jersey, 08544, USA}
\date{\today}

\begin{abstract}
The deformation and rupture of a lipid vesicle due to the forced normal approach of an inclusion are essential for optimizing the design of magnetic giant unilamellar vesicles [magGUVs, Malik et al., Nanoscale 17, 13720 (2025)], with implications for active colloid-membrane interactions and cellular-scale chemical delivery. Here, we investigate vesicles propelled by a force-driven rigid inclusion and reveal a robust elastohydrodynamic mechanism: the inclusion outpaces the vesicle, sustaining a thinning film that drains symmetrically and self-similarly, largely independent of initial shape. For soft membranes and small inclusions, coupling drives a monotonic tension increase that can exceed the lysis tension. Evaluating the maximal tension over a delivery distance, we map an operating window in vesicle reduced area and size relative to the inclusion.
\end{abstract}

\pacs{Valid PACS appear here}
                              
\maketitle


Lipid vesicles encapsulating colloids or microswimmers provide an \emph{in vitro} platform for active-membrane physics, with membrane–inclusion interactions setting the dynamics relevant to protocell evolution~\cite{blain2014ARB,babu2022motile,miele2022shape,minagawa2025self}, pathogen motility~\cite{iye-gom-fed2022}, and targeted delivery~\cite{bulbake2017liposomal,Liu2022Molecules}.
%
%
Experiments with encapsulated \emph{Bacillus subtilis} and active Janus particles reveal vigorous reshaping of vesicles without sustained directed motion~\cite{takatori2020active,willems2025run}, whereas vesicles enclosing \emph{Escherichia coli} display persistent propulsion via mechanical coupling between the flagellar bundle and the membrane tethers, forming a rotating helical assembly that drives the vesicle~\cite{nag-bro-daw-mar-poo-sta2022}. Both theory and computation link swimmer-induced vesicle shape changes to its propulsion~\cite{pao-dil-mar-ang2016,chen2017rotational,reigh2017swimming,dad-goh-lie-hoe-mat-guz-sch-men-low2019,wang2019shape,shan2019assembly,sprenger2020towards,vutukuri2020active,young2021many,ruske2021morphology,marshall2021hydrodynamics,kree2021controlled,iyer2023dynamic}, underscoring that, whether actively or passively driven, propulsion is set by how forces and torques are transmitted through the \emph{soft-lubrication} film that separates the inclusion from the membrane~\cite{xiao2022force,skotheim2004soft,bertin2022soft,rallabandi2024fluid,fessler2025energetics,malik2025magnetically}.


This soft-lubrication framework also underpins collisions between a rigid
sphere and a fluid interface or a thick elastic plate in viscous media,
and connects to Hertzian-type quasistatic impacts with elastic
sheets~\cite{gopinath2011elastohydrodynamics,ral-opp-zio-sto2018, daddi2018reciprocal, far2025}, where a balance between elastic, viscous, and inertial dissipations yields similarity laws for the
film thickness and pressure~\cite{snoeijer2013similarity}. Classic work
by \citet{jones1978film} demonstrated self-similar thinning for a sphere
settling toward a fluid--fluid interface, with scalings set by the ratio
of external forcing to surface tension: near a rigid wall, tension
dominates forcing, and an intervening thin liquid film drains exponentially, $h\sim e^{- t/\tau}$ (with $\tau$ a time constant~\cite{stone2005lubrication}), while for a deformable interface of finite tension one finds $h\sim t^{-1/4}$ (weak forcing) and $h\sim t^{-1/2}$ (moderate forcing)~\cite{jones1978film}.

\begin{figure}[h]
  \centering
  \includegraphics[width = 0.45\textwidth]{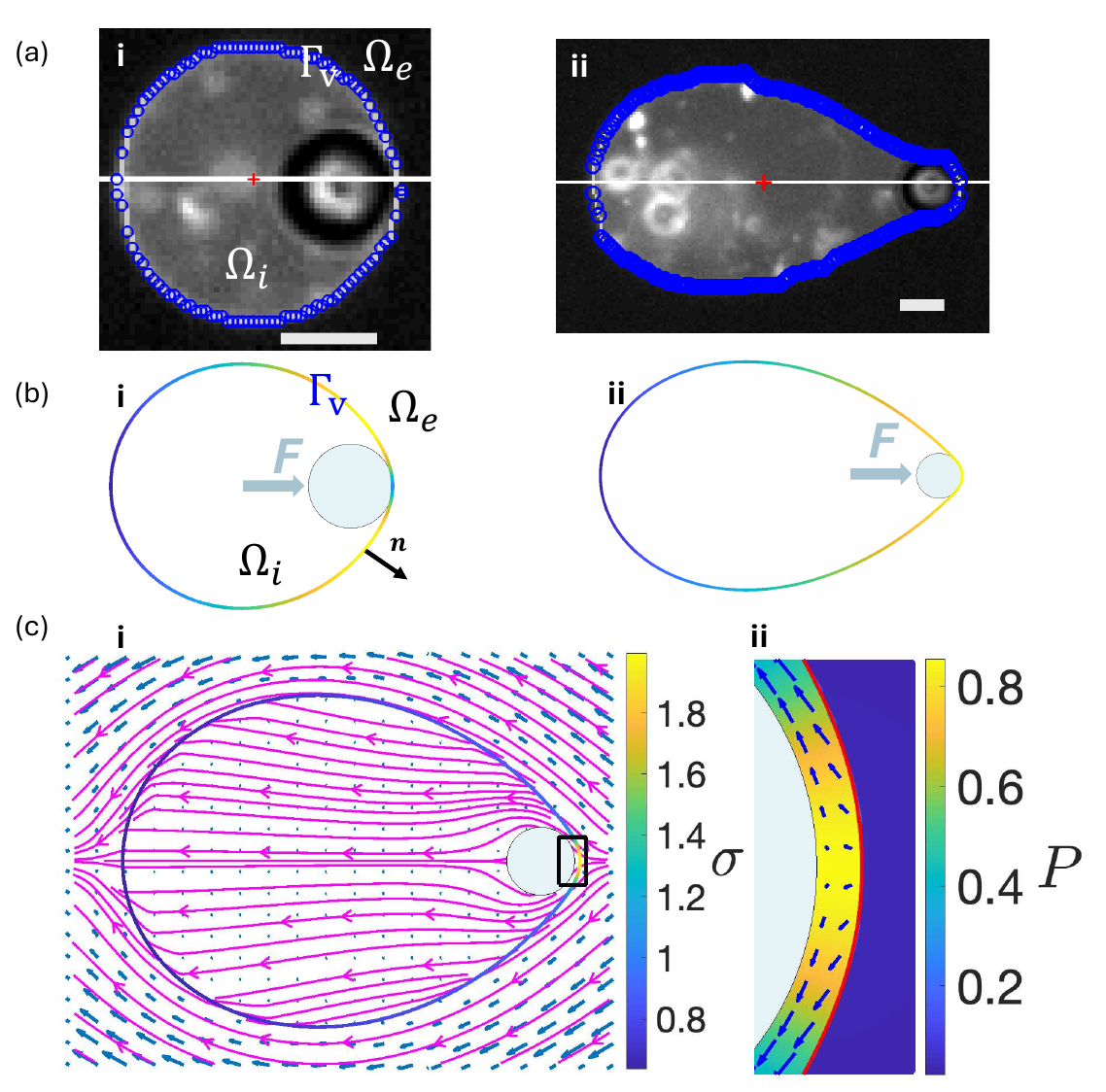}
  \caption{\label{figure1} A vesicle pushed by an inclusion particle under a constant force deforms and translates. (a) A magGUV under a constant, uniform magnetic field gradient. Scale bars are $10$ $\mu$m. (b) Quasi-steady simulated equilibrium shapes from 2D boundary-integral simulations, which produce the (c) flow field around a vesicle pushed by a rigid inclusion (blue circle). The vesicle membrane is color-coded (i) by its tension $\sigma$, and the draining flow in the interstitial space is driven by the pressure gradient (ii).}
\end{figure}

We investigate soft-lubrication flow generated as a rigid inclusion approaches a vesicle modeled as an inextensible fluid membrane with spatially varying tension. Lubrication flows driven by tangential motion near compliant substrates are well documented~\cite{bureau2023lift,rallabandi2024fluid}, but we focus on the \emph{normal} approach toward a membrane of constant bending rigidity and nonuniform tension, enforcing inextensibility. As in the confinement of a squirmer within a viscous drop~\cite{reigh2017swimming}, hydrodynamic coupling between a force-driven colloid and the membrane produces vesicle propulsion. Experiments across colloid sizes and vesicle deformabilities---from a nearly rigid spherical vesicle [Fig.~\ref{figure1}(a)i] to a deformable one [Fig.~\ref{figure1}(a)ii]---show co-translation of the inclusion and vesicle at nearly identical speeds with an accompanying steady vesicle shape~\cite{malik2025magnetically}. 
Guided by these results, we focus on the soft-lubrication drainage of the interstitial film---a regime that is difficult to probe experimentally and challenging to simulate using generic stencil-based CFD or standard lattice-Boltzmann solvers.

We start from the incompressible Stokes equations for fluid at $\xx \in \Omega=\Omega_e\cup \Omega_i$ [Fig.~\ref{figure1}]
\begin{align}\label{eq:Stokes} 
\nabla\cdot{\boldsymbol{\tau}}=- \nabla P+\mu \Delta \uu = {\bf 0}, 
    \qquad \nabla \cdot \uu = 0,
\end{align}
where ${\boldsymbol{\tau}} = -P{\boldsymbol{I}} + \mu( \nabla \uu + \nabla \uu^T)$ is
the stress tensor with pressure $P$, the incompressible flow field $\uu$, and the fluid viscosity $\mu$ (assumed to be identical in both $\Omega_i$ and $\Omega_e$). The
hydrodynamic force density on the vesicle membrane is balanced with internal and external forces: $\llbracket {\boldsymbol{\tau}} \rrbracket \cdot \nn = \ff$ for $\xx \in \Gamma_{\text{v}}$,
where $\llbracket \cdot \rrbracket$ denotes the jump across
$\Gamma_{\text{v}}$, $\nn$ is the unit outward normal on
$\Gamma_{\text{v}}$, and the membrane force density $\ff$ is the sum of bending and membrane tension stresses. We use the magnitude of the external force per length (in 2D) $|\FF|$ on the inclusion, the inclusion radius $a_0$, and viscosity $\mu$ to non-dimensionalize the equations (see Supplemental Material~\cite{supplementary}). The vesicle-inclusion system is then parametrized by a dimensionless bending stiffness $\kappa_b$ (membrane bending stiffness scaled to $|\FF|a_0^2$), a size ratio $\bar{b}=\sqrt{A_0/(\pi a_0^2)}$, and membrane deformability, which in 2D can be characterized by the reduced area $\nu=4\pi A_0/L_0^2$ with $A_0$ the enclosed area and $L_0$ the circumference of the membrane. We developed a fast, high‐order 2D boundary‐integral solver to simulate the coupled fluid–membrane dynamics (\cite{quaife2014high}, see Supplemental Material~\cite{supplementary}). The method efficiently resolves the lubrication flow in the near‐contact gap between the inclusion and membrane with high fidelity [Fig.~\ref{figure1}(b,c)], and the scaling analysis below shows that the resulting similarity structure is identical between 2D and 3D axisymmetric. 
By contrast, fully 3D axisymmetric simulations that resolve near-contact soft-lubrication are far more expensive: they require high-order surface discretizations, near-singular quadrature, and much smaller time steps~\cite{vee-gue-bir-zor2009b}.
An alternative based on the method of fundamental solutions can capture lubrication between nearly touching \emph{rigid} surfaces~\cite{bro-bar-tor2025}, but it becomes severely ill-conditioned as the gap closes and is inapplicable when one boundary is deformable.

\begin{figure}[h!]
 \centering
  \includegraphics[width = 0.45\textwidth]{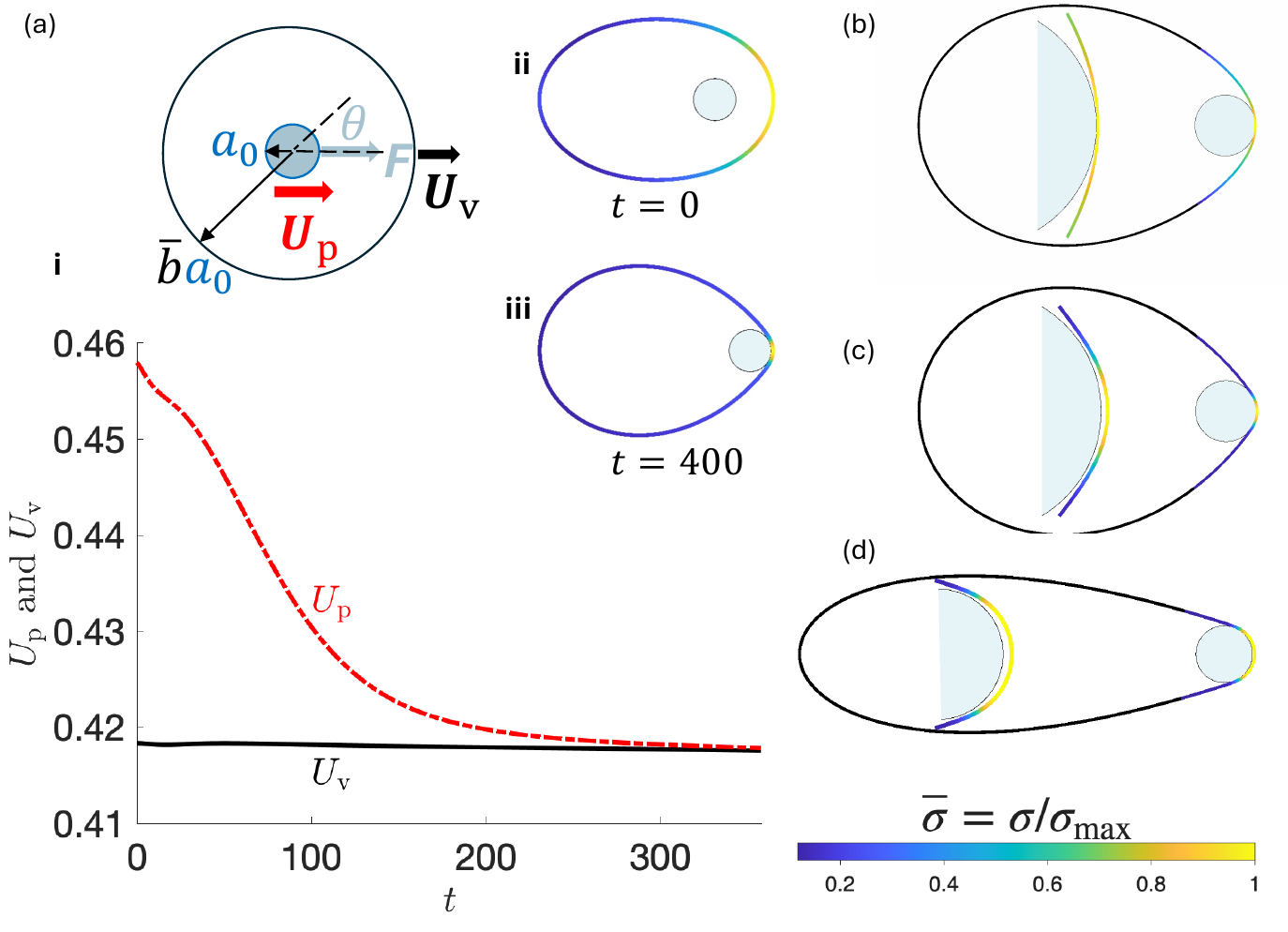}
\caption{\label{figure2} (a) Transient dynamics to the quasi-steady equilibrium configurations, and (b,c,d) quasi-steady equilibrium shapes color-coded by $\bar\sigma=\sigma/\sigma_{\mathrm{max}}$. (a, i) Particle speed $U_{\mathrm{p}}$ and vesicle speed $U_{\mathrm{v}}$ as a function of time. Schematic showing the polar angle $\theta$ with respect to the forcing direction. 
(a, ii\&iii) Configurations at $t=0$ and $t=400$. (b) A stiff membrane ($\kappa_b=1$ and $\nu=0.95$) and a quadratic film profile in the inset. (c) A soft, less deformable membrane ($\kappa_b=10^{-3}$ and $\nu=0.95$) and a dimple-shaped film profile in the inset. (d) A soft, more deformable membrane ($\kappa_b=10^{-3}$ and $\nu=0.65$) and a dimple-shaped film profile in the inset.}
\end{figure}

Our 2D simulations at matched parameters reproduce the experimentally measured trends [Fig.~\ref{figure1}(b, i\&ii)]: in the near‐contact regime, lubrication pressures dominate, drive finite membrane deformations and generate strong spatial tension $\sigma$ gradients that regulate a nonlinear response [Fig.~\ref{figure1}(c, i\&ii)]. Within the interstitial film, the pressure peaks on the symmetry axis, regulating the drainage flow that controls the overall approach of the inclusion toward the membrane.
%
%
Simulations show that, under constant forcing on the rigid inclusion, the vesicle (speed $U_{\mathrm{v}}$) always translates slower than the inclusion (speed $U_{\mathrm{p}}$)---even in near-contact when the intervening lubrication film drains [Fig.~\ref{figure2}(a, i)]. From the initial configuration at $t=0$ [Fig.~\ref{figure2}(a, ii)], the inclusion initially outruns the membrane, approaches near-contact, and then the pair co-translate in a quasi-steady equilibrium by 
$t=400$ [Fig.~\ref{figure2}(a, iii)].

The same transient dynamics occurs for all initial conditions. For a spherical inclusion in 3D (circular inclusion in 2D), the forcing direction ($\theta=0$ in Fig.~\ref{figure2}(a,i)) sets the symmetry axis and the quasi-steady equilibrium shape. Three robust morphologies emerge: (i) a \emph{stiff} membrane with nearly uniform tension forms a locally quadratic film [Fig.~\ref{figure2}(b)]; (ii) a \emph{soft, less deformable} membrane produces a dimple with a pronounced tension peak at the symmetry axis [Fig.~\ref{figure2}(c)]; and (iii) a \emph{soft, more deformable} membrane develops a dimple with an extended neck and nearly uniform tension [Fig.~\ref{figure2}(d)].


Once near-contact is reached, the subsequent dynamics are insensitive to the initial state: the external forcing sets the symmetry axis, the membrane relaxes to a quasi-steady shape, and the lubrication layer thins monotonically. For an axisymmetric inclusion, the film profile $h(\theta, t)$ is a function of the polar angle $\theta$ (with the on-axis film thickness denoted by $h_0(t)\equiv h(\theta=0,t)$). 
The mass conservation of fluid in the interstitial space between an elastic, inextensible membrane (hence a non-uniform tension $\sigma(\theta,t)$) and a normally approaching rigid inclusion yields the dimensionless lubrication equation 
\begin{equation}
\label{eq:lub_eq00}
   \partial_t h =\frac{1}{g(\theta)}\partial_{\theta}\left[g(\theta) \left(\frac{\epsilon^2}{3}\left(h^3\partial_{\theta}P\right)-\frac{\epsilon}{2}\left(h^2\partial_{\theta}\sigma\right)\right)\right],
\end{equation}
where $\epsilon\ll 1$ is the aspect ratio of film height to particle radius, and $g(\theta)=1$ and $g(\theta)=\sin\theta$ for 2D and 3D, respectively. The pressure $P$ in the thin film is determined by both the membrane shape and the membrane tension determined from the area incompressibility. We provide a local analysis of Eq.~\eqref{eq:lub_eq00} in the Supplemental Material~\cite{supplementary} to illustrate the existence of both a quasi-steady quadratic film profile and a dimple-shaped film profile.

\begin{figure}
 \centering
  \includegraphics[width = 0.5\textwidth]{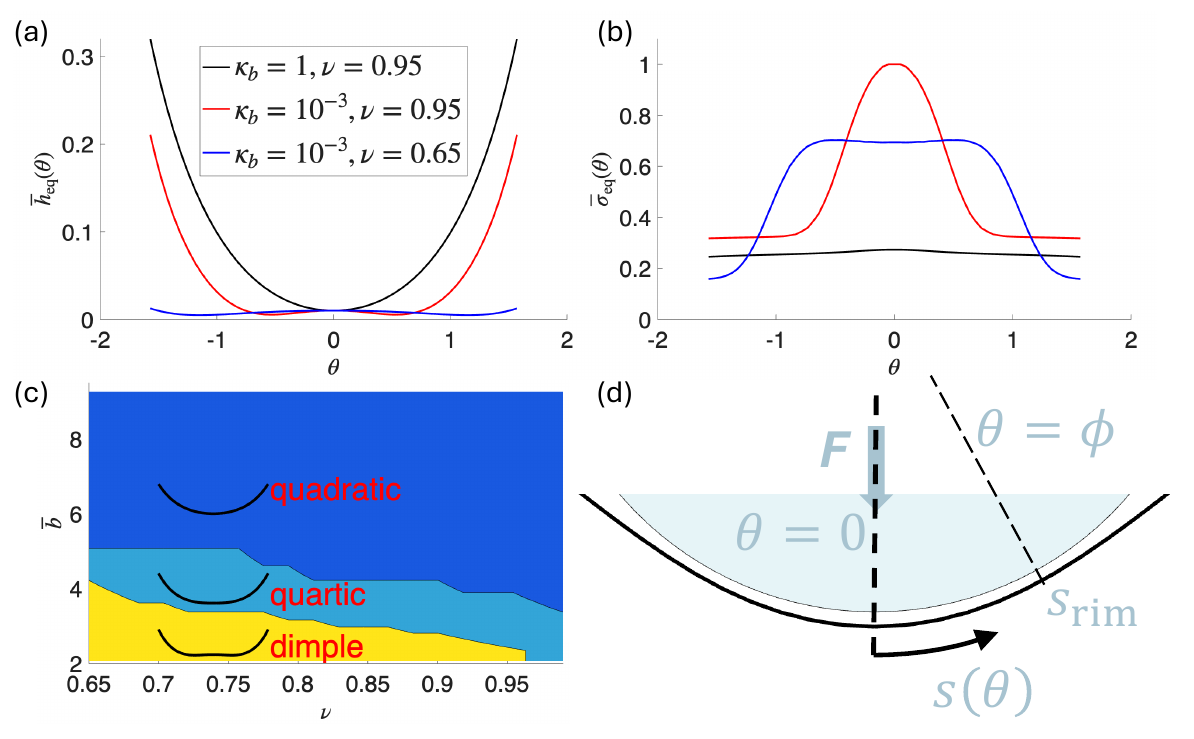}
\caption{\label{figure3} Quasi-steady thin film characteristics. (a) 2D quasi-steady equilibrium film profiles, and (b) their corresponding tension distributions. Black curves: stiff vesicle; red curves: soft and less deformable vesicle; blue curves: soft and more deformable vesicle. (c) Distribution of equilibrium film profiles in size ratio $\bar{b}$ and reduced area $\nu$ for a stiff membrane ($\kappa_b = 1$). (d) Dimple-shaped film profile with arc-length as a function of the angle $s(\theta)$ around the neck at the rim $\theta=\phi$.}
\end{figure}

First we focus on the quasi-steady shapes ($\partial_t h \approx 0$).
In 2D, three robust morphologies emerge, organized by bending stiffness $\kappa_b$ and reduced area $\nu$.
(i) For a \emph{stiff, less-deformable} membrane ($\kappa_b \sim 1$, $\nu$ near unity), the film is locally quadratic with nearly uniform tension [black; Fig.~\ref{figure3}(a,b)].
(ii) For a \emph{soft, less-deformable} membrane ($\kappa_b \ll 1$, $\nu$ near unity), the profile is \emph{dimpled} and the tension exhibits a sharp peak on the symmetry axis [red; Fig.~\ref{figure3}(a,b)].
(iii) For a \emph{soft, more-deformable} membrane ($\kappa_b \ll 1$, smaller $\nu$), the dimple develops an \emph{extended neck} with a broadly distributed tension across that region [blue; Fig.~\ref{figure3}(a,b,d)].
To close the quasi-steady profiles, the tension field required by local area-incompressibility is obtained from the 2D simulations.
A morphology map for $\kappa_b=1$ in the $(\bar b,\nu)$ plane is shown in Fig.~\ref{figure3}(c); for $\kappa_b=10^{-3}$, the quasi-steady film is dimpled within the explored range $\nu\in[0.65,\,0.95]$.


Local analysis on the quasi-steady equilibrium 3D axisymmetric thin film shows that the quadratic profile is possible only when the squeezing flow is between two rigid boundaries, such as a spherical inclusion against a spherical vesicle. Furthermore, we find that the thin-film profile between a rigid sphere squeezing against a deformable membrane is always dimple-shaped, and the convexity depends on the Marangoni curvature (see Supplemental Material \cite{supplementary}).
%
The dimple-shaped thin-film profile in Fig.~\ref{figure3}(a) is similar to the liquid film between an elastic sheet and a solid wall~\cite{young2014long,young2017long}, between adhesive vesicles~\cite{quaife2019hydrodynamics}, and between viscous drops in squeeze-flow-like configurations under force~\cite{bai2021water,jones1978film} or shear~\cite{tsekov1994dimple,zhou2002numerical,lowengrub2016numerical}, where soft-lubrication flow drives self-similar drainage as the minimum film thickness tends to zero.

Next we focus on the draining of a dimple-shaped lubrication film as a rigid inclusion approaches an elastic inextensible membrane. The drainage of a quadratic film profile is related to the thinning film between a rigid sphere and a rigid wall, where the self-similar scaling is drastically different between 2D and 3D \cite{stone2005lubrication}. Our matched asymptotic analysis of the inner region around the film neck (Fig.~\ref{figure3}(d); see Supplemental Material) shows that the drainage scaling of a dimple-shaped film is universal between 2D and 3D axisymmetric geometries; our 2D boundary-integral simulations uncover the self-similar scalings that depend on both membrane stiffness $(\kappa_b)$ and deformability $(\nu)$ as we elucidate below. 
\begin{figure*}[t]
  \centering
\includegraphics[width=\textwidth]{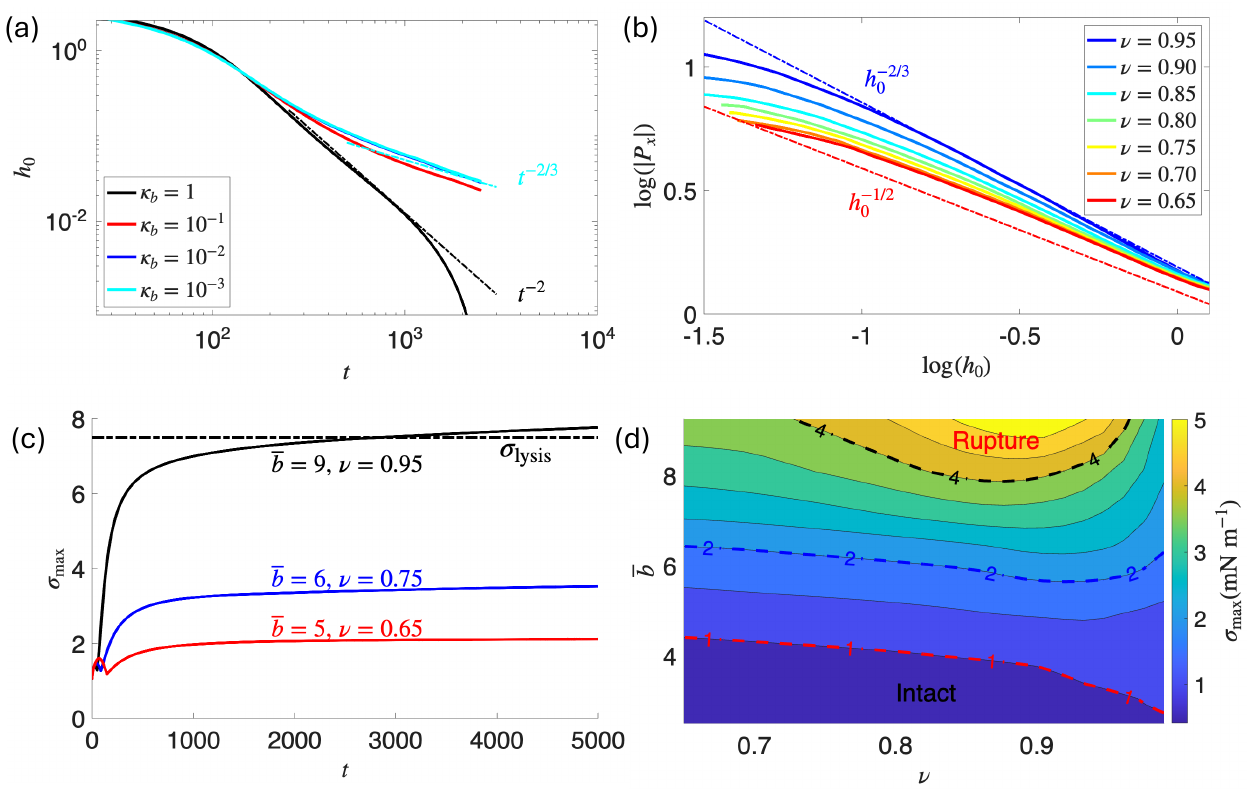}
  \caption{Scaling of thin-film drainage from simulations.
(a) Neck drainage for \(\nu=0.95\): \(h_0(t)\) decays self-similarly.
(b) Pressure gradient–height scaling (\(|P_x|\) vs \(h_0\)) for a soft membrane (\(\kappa_b=10^{-3}\)) as deformability varies from \(\nu=0.95\) (blue) to \(\nu=0.65\) (red).
(c) Maximum membrane tension \(\sigma_{\max}\) for three pairs of size ratio \(\bar b\) and $\nu$ at \(\kappa_b=10^{-3}\); dashed line: lysis tension \(\sigma_{\mathrm{lysis}}\) scaled to \(|\mathbf F|=5\times10^{-4}\,\mathrm{N\,m^{-1}}\).
(d) Map of dimensional \(\sigma_{\max}\) (mN\,m\(^{-1}\)) over \((\nu,\bar b)\); the black dashed contour is \(\sigma_{\max}=\sigma_{\mathrm{lysis}}=4\,\mathrm{mN\,m^{-1}}\). Rupture is predicted above this contour.
\label{figure4}}
\end{figure*}
In particular, simulations of a rigid disk approaching a deformable, inextensible membrane reveal distinct drainage regimes [Fig.~\ref{figure4}(a,b)]: the draining follows $h_0(t)\sim t^{-\alpha}$ with $\alpha$ set by the bending stiffness $\kappa_b$ and deformability $\nu$. For a \emph{stiff} membrane $(\kappa_b=1)$ we recover $\alpha \approx 2$, matching the rigid-wall limit of a cylinder approaching a plane~\cite{stone2005lubrication} across a range of values for $\nu \in [0.65,0.99]$. For a \emph{soft} membrane ($\kappa_b=10^{-3}$) we numerically find $\alpha \approx 2/3$ over the same range of $\nu$, larger than the fluid–interface case of a rigid cylinder~\cite{jones1978film}, reflecting the finite bending resistance that shifts the dynamics toward the rigid-wall limit. 
The pressure gradient scales with the neck height as \(|P_x|\sim
h_0^{\,\beta}\), with \(\beta\) varying with deformability:
\(\beta\approx -2/3\) at \(\nu=0.95\) and \(\beta\approx -1/2\) at
\(\nu=0.65\), while the drainage exponent remains fixed at
\(\alpha\approx 2/3\) [Fig.~\ref{figure4}(b)]. Under constant forcing, the drainage of a quadratic film profile is fixed by a single pair $(\alpha,\beta)$. By contrast, dimpled profiles exhibit non-unique $\beta$
at the same $\alpha$, indicating \emph{two} tangential length scales—one governing pressure variation, the other advective outflow—consistent with the scaling analysis in the Supplemental Material.


To understand such complex interplays between multiple scales in the drainage dynamics of a dimpled-shaped thin film, we focus on the drainage in the ``inner" region where the film height is smallest at the dimple neck~\cite{jones1978film}. We introduce inner similarity variables along the membrane arclength $s$ measured from the rim at $\theta=\phi$ [Fig.~\ref{figure3}(d)]: $s-s_{\rm rim}(t')=\ell(t')\xi$,
$h(s,t')=H(t')J(\xi)$, $\sigma(s,t')=\sigma_0(t')G(\xi,t')$, with general power-law scalings in term of the amplitude $j=j(t')$ that encode the self-similar behavior: $H=j^{2\hat\alpha},\, \ell=\hat\beta\,j^{\hat\lambda},\, \sigma_0=j^{\hat c_0},\,
G=j^{\hat c_1}\widetilde G(\xi).$
Here $t'$ is the characteristic time for inner scaling~\cite{jones1978film}, and $J(\xi)$ and $\widetilde G(\xi)$ are order-one similarity shapes.
%
%
%
The equation for the depth-averaged interfacial flux, 
under small-slope linearization (capillarity and bending) in the inner region of Eq.~\eqref{eq:lub_eq00}, is
\begin{equation}
\label{eq:InnerDraining}
\partial_{\xi}^3J -\frac{\kappa_b}{\hat\beta^2}j^{-(2\hat\lambda+\hat c_0)}\partial_{\xi}^5J - \frac{3\hat\beta^2}{2J} j^{2\hat\lambda-4\hat\alpha+\hat c_1}\partial_{\xi}\widetilde G
= 
{\cal F} \frac{d j}{d t'},
\end{equation}
where ${\cal F} = j^{\,3\hat\lambda-8\hat\alpha}/(g(\phi)J^3)$,
with 
$g(\phi)=1\,(\sin\phi)$ for the 2D cylinder (3D axisymmetric) geometry. To leading order, the 2D and 3D problems share the \emph{same} similarity operator and exponents. Therefore, the drainage scaling in the 2D simulations shed light on the self-similar draining dynamics that is formidable in 3D axisymmetric boundary-integral simulations.

Next, for self-similar drainage, we require the powers of $j(t')$ multiplying the operator terms in Eq.~\eqref{eq:InnerDraining} to vanish to obtain a genuine $\xi$-ODE whose coefficients do not depend on $t'$. This yields the separation conditions: 
\begin{align}
2\hat\lambda+\hat c_0=0 
\text{  and  } \ 2\hat\lambda-4\hat\alpha+\hat c_1=0.
\label{eq:sep_conditions}
\end{align}
The \emph{forcing} exponents due to squeeze drainage and Marangoni stress are then fixed by the measured drainage and gradient scalings from simulation. Next, we combine the separation conditions 
in Eq.~\eqref{eq:sep_conditions} 
with the numerically determined scaling in Fig.~\ref{figure4}(a,b) for the height decay and for the pressure-gradient and tension-gradient scalings to solve for $(\hat\alpha,\hat\lambda,\hat c_0,\hat c_1)$ for two reduced areas: $\nu=0.65$ (red solid curve in Fig.~\ref{figure4}(b), highly deformable) and $\nu=0.95$ (blue solid curve in Fig.~\ref{figure4}(b), less deformable). For $\nu=0.65$, the drainage scaling in Fig.~\ref{figure4}(a) and the gradient scaling in Fig.~\ref{figure4}(b) are 
\begin{align}\label{eq:inner_odes_case1}
\frac{2}{\,3\hat\lambda-8\hat\alpha+1\,} =-\frac{2}{3}, \;\;\;
\frac{\hat c_0+\hat c_1-\hat\lambda}{2\hat\alpha} =-\frac{1}{2}.
\end{align}
Together with Eq.~\eqref{eq:sep_conditions}
we obtain
$\hat\alpha=\hat\lambda=4/5$ and $\hat c_0=-\hat c_1=8/5$.
Hence the membrane tension scales as $\sigma\sim \sigma_0 G\sim j^{\hat c_0+\hat c_1}\sim H^{(\hat c_0+\hat c_1)/(2\hat\alpha)}=H^0$: it asymptotes to a constant during drainage when the membrane is soft and highly deformable (red and blue curves in Fig.~\ref{figure4}(c)).
For $\nu=0.95$, the drainage and gradient scalings imply
\begin{align}\label{eq:inner_odes_case2}
\frac{2}{\,3\hat\lambda-8\hat\alpha+1\,} =-\frac{2}{3},\;\;\;
\frac{\hat c_0+\hat c_1-\hat\lambda}{2\hat\alpha} =-\frac{2}{3},
\end{align}
which corresponds to 
$\hat\alpha=5/6$, $\hat\lambda=8/9$,
$\hat c_0=-5/3$, and
$\hat c_1=14/9$.
Therefore $(\hat c_0+\hat c_1)/(2\hat\alpha)=-1/15$ and $\sigma \sim H^{-1/15}$, indicating a weak divergence of tension as the film drains when the membrane is soft but less deformable, as shown by the black curve in Fig.~\ref{figure4}(c).


In \citet{malik2025magnetically}, the vesicle–inclusion compound particle remains intact during transport, and ruptures only upon photoactivation at the target: magGUV with an inclusion radius $a_0 = 3$ $\mu$m at a speed of $\sim 60~\mu\mathrm{m\,s^{-1}}$ can travel for $\sim 60~\mathrm{s}$ without rupture. This is consistent with our soft-lubrication analysis: as the inclusion approaches, the membrane tension increases while the minimal gap drains self-similarly. It also reveals a limitation—under sustained drive the tension can exceed the lysis tension (between 3 to 4 mN m$^{-1}$ for a GUV under slow loading \cite{portet2010new}) before arrival. Using a 2~mm travel distance as a benchmark, 2D simulations with \(a_0=3~\mu\mathrm{m}\) and $|\mathbf F|\sim 10^{-8}~\mathrm{J\,m^{-2}}$ ($|\mathbf F|a_0^2=10^{-19}~\mathrm{J}$) keep peak tension below the lysis threshold (blue and red curves in Fig.~\ref{figure4}(c)), whereas $a_0=0.5~\mu\mathrm{m}$ and \(|\mathbf F|\sim5\times10^{-4}~\mathrm{J\,m^{-2}}\) $(|\mathbf F|a_0^2=10^{-16}~\mathrm{J}$) exceed the lysis tension before \(2~\mathrm{mm}\) (black curve). Fig.~\ref{figure4}(d) maps the maximum tension over the travel distance versus size ratio $\overline b$ and reduced area $\nu$, delineating rupture regions that we expect to be universal for both 2D and axisymmetric 3D vesicle membranes.

We examined the normal-approach elastohydrodynamics of a rigid inclusion driven toward a deformable vesicle. Membrane bending and inextensibility organize the expelled soft-lubrication flow into a robust similarity regime: the inclusion persistently outpaces the vesicle, the interstitial film height drains self-similarly ($h_0\!\sim\!t^{-2/3}$ for soft membranes), and the membrane tension rises monotonically—approaching a finite plateau for highly deformable vesicles and growing toward rupture for less deformable ones.  These results allow us to estimate an upper bound on the time to lysis. In the near-contact limit, inclusion surface properties (roughness, short-ranged adhesion/repulsion) are expected to shift drainage and rupture conditions, suggesting surface-design routes for controllable release in magGUVs and related delivery platforms~\cite{xiao2022force,wan2024nonequilibrium,fessler2025energetics}.
This picture parallels elastohydrodynamic adhesion in soft solids—where an \emph{apparent-contact radius} organizes early sticking 
and late peeling---while in our membrane system, the \emph{neck} plays the analogous role, with similarity-controlled drainage governed by two tangential scales 
(pressure variation and advective outflow) 
that select the observed exponents~\cite{bertin2025sticking}.
Augmenting linear elasticity by Helfrich bending with inextensible tension yields a deformability-dependent tension–height law that selects neck shapes (quadratic versus dimple) and scalings.
Together, these results articulate a ``membrane-elastohydrodynamics’’ mechanism---balancing curvature-pressure (bending/variable tension) with viscous drainage at the neck---that organizes regimes and offers a direct handle to avoid premature rupture or to enable on-demand release.

\acknowledgments
Y.N.Y.~acknowledges support from NSF (DMS-1951600 and DMS-2510714) and Flatiron Institute, part of Simons Foundation. B.Q.~acknowledges support from NSF (DMS-2510713). H.N.~acknowledges support from NSF (DMS-2211633).
O.S.P.~acknowledges partial support from NSF (CBET-2323046 and CBET-2419945).
J.F.~acknowledges partial support from NSF (CBET-2323045).
H.A.S.~acknowledges support from NSF (CBET-224679). 

\bibliography{refs}

\end{document}